\documentstyle[psfig,times]{mn}
\newif\ifAMStwofonts
\newcommand{\gapp}{\mbox{\raisebox{-0.3em}{$\stackrel{\textstyle >}{\sim}$}}}

\title[The dynamics of the giant radio galaxy 3C\,457]{The dynamics of the giant radio galaxy 3C\,457}
\author[C. Konar et al.]
       {C. Konar$^{1}$ $\thanks{E-mail: chiranjib@iucaa.ernet.in (CK), 
       m.j.hardcastle@herts.ac.uk (MJH), J.Croston@soton.ac.uk (JHC), djs@ncra.tifr.res.in (DJS) }$, 
       M.J. Hardcastle$^{2}$, J.H. Croston$^{3,2}$ and D.J. Saikia$^{4,5}$  \\
$^{1}$ Inter-University Centre for Astronomy and Astrophysics, Pune University Campus, Post Bag 4,
Pune 411 007, India \\
$^{2}$ School of Physics, Astronomy \& Mathematics, University of Hertfordshire, College Lane,
       Hatfield AL10 9AB  \\ 
$^{3}$ School of Physics and Astronomy, University of Southampton, Southampton SO17 1BJ \\
$^{4}$ National Centre for Radio Astrophysics, TIFR, Pune University Campus, Post Bag 3,
Pune 411 007, India \\
$^{5}$ Australia Telescope National Facility, CSIRO, PO Box 76, Epping, NSW 1710, Australia \\
}
\date{Accepted.    Received }

\pagerange{\pageref{firstpage}--\pageref{lastpage}}
\pubyear{  }

\begin{document}

\maketitle

\label{firstpage}

\begin{abstract}
We present multi-frequency radio observations with the Giant Metrewave
Radio Telescope and Very Large Array, and X-ray observations with the
X-ray Multi-Mirror Mission ({\it XMM-Newton}) telescope of the giant radio
source (GRS) 3C\,457. We have detected the core, lobes and the
environment of the GRS in X-ray. We examine the relationships between
the radio and X-ray emission, determine the radio spectrum over a
large frequency range and attribute the X-ray emission from the lobes
to the inverse-Compton scattering of cosmic microwave background (CMB)
photons. The magnetic field strength of the lobes is very close to the
equipartition value. Both the lobes are in pressure balance near the
hotspots and apparently under-pressured towards the core. The X-ray
spectrum of the core of the GRS consists of an unabsorbed soft
power-law component and a heavily absorbed hard power-law component.
The soft unabsorbed component is likely to be related to the radio
jets. There is no strong evidence of Fe K$\alpha$ emission line in our
data.

\end{abstract}


\section{Introduction}
Giant radio sources (GRSs) are defined to be those which have a
projected linear size $\gapp$1 Mpc (H$_o$=71 km s$^{-1}$ Mpc$^{-1}$,
$\Omega_m$=0.27, $\Omega_{vac}$=0.73, Spergel et al. 2003). The lobes
of GRSs extend well beyond the interstellar medium (ISM) and the
coronal halos of the host galaxies. Therefore the light synchrotron
plasma interacts with the heavier intergalactic medium (IGM). The very
existence of GRSs raises important questions about the properties of
the IGM they inhabit. For radio lobes to exist, there must be a medium
to confine them, either through thermal pressure or ram pressure. On
Mpc scales, the external pressure even in a rich cluster must be
several orders of magnitude lower than in the typical environments of
smaller radio galaxies, and in fact we know already that GRSs do not
inhabit particularly rich environments. What phase of the IGM governs
the dynamics of giant radio sources? If we could answer this question,
it would be possible to use GRSs at cosmological redshifts to probe
the evolution of this phase over cosmic time. Studying the GRSs, we
might also be able to determine whether these are simply probing the
extreme edge of the intra-group or intra-cluster media, or whether
they require some other medium, such as the Warm Hot Intergalactic
Medium (WHIM), for lobe formation on these scales to be possible. In a
recent piece of work Safouris et al. (2008) have shown that the GRS
MRC B0319-454 appears to be embedded within a large scale galaxy
filament. Croston et al. (2004) reported that the lobes of large radio
sources are close to pressure balance near the hotspots and are
under-pressured towards the core, if it is assumed that the protons
are not energetically dominant in radio lobes as is likely to be the
case (Croston et al. 2005). This means that the over-pressured lobes
required by self similar models (Falle 1991; Kaiser \& Alexander 1997;
Kaiser, Dennett-Thorpe \& Alexander 1997) of FRII radio sources are no
longer supported by observations, at least for large radio sources, as
discussed by Hardcastle \& Worrall (2000). This raises important
questions about how such large radio galaxies evolve.

To resolve the various issues raised in the above paragraph it is
necessary to measure the internal and external pressure of the lobes
of giant radio sources. If the GRS environment consists of a hot group
or cluster scale hot-gas medium then the external pressure can be
measured via X-ray observations of thermal bremsstrahlung emission
from the intracluster medium (ICM). X-ray observations (combined with
radio ones) also allow us to measure internal pressure of the lobes
via the inverse-Compton (IC) process, in which the relativistic
electrons in the lobe scatter photons from the cosmic microwave
background (CMB), (hereafter IC-CMB). Detections of IC emission have
the potential to clarify the particle content and magnetic field
strength of radio galaxies because they allow direct measurements of
the electron energy density, unlike observations of radio synchrotron
emission where the electron density and the magnetic field strength
cannot be decoupled (cf. Croston et al. 2005). This technique has been
used successfully to estimate magnetic field strengths in the lobes
and hotspots of many smaller Fanaroff-Riley class II (Fanaroff \&
Riley 1974) sources, and constrain source dynamics and particle
content by comparing the internal pressure with the external pressure
from X-ray emitting hot gas (Hardcastle et al. 2002; Croston et al.
2004). An extensive study of X-ray emission from the lobes of a sample
of FRII radio galaxies and quasars has shown that although a few may
be magnetically dominated by factors of 2 or more, about 70 per cent
of the sample have magnetic field strengths within $\sim$35 per cent
of the equipartition value, or electron dominance $(U_e/U_B)$ by a
factor of $\sim$5 (Croston et al. 2005; see also Kataoka \& Stawarz
2005). The key point, however, is that observations of inverse-Compton
radiation allow us to measure the internal pressure in the lobes of
individual objects, which then allows us to understand the dynamics of
the radio source.

To date there has been little attempt to study the lobes of powerful
GRSs using this technique. Partly this is a consequence of much of the
early work having been done with {\it Chandra}, with its relatively
small field of view and poorer surface brightness sensitivity.
However, the work of Croston et al. (2004), who used {\it XMM-Newton}
to observe the low-redshift, 600$-$700 kpc sources 3C\,223 and
3C\,284, shows the potential of such studies. Inverse-Compton
detections of true giant sources are essential if we are to measure
the lobe pressures, and thus the pressure of the confining medium, in
this poorly understood population.

The giant radio sources with very extended diffuse radio lobes are
very suitable for such kind of analysis, as (for a given radio
luminosity) larger sources produce more IC-CMB emission. We have
chosen to look for IC emission for the first time from a GRS, 3C\,457,
so as to investigate various issues, including pressure balance of the
radio lobes, equipartition between magnetic field and particles in
lobes, and the interaction of the lobes and the external environment.
3C\,457 is an FRII-type (Fanaroff \& Riley 1974) giant radio galaxy with
a largest angular size of 190 arcsec. It is situated at a
redshift of 0.428 (Perryman et al. 1984). For our cosmology, defined
in the beginning of this section, the redshift of 0.428 corresponds to
a scale of 5.567 kpc arcsec$^{-1}$ giving a largest linear size of
1058 kpc. 3C\,457 has two large, symmetric diffuse lobes. It is
comparatively bright at radio wavelengths and so it is likely to be
detected in X-rays via IC-CMB emission. Moreover, since it is a radio
galaxy, not a quasar, it is easier to separate the nuclear emission
from the lobe-related X-ray emission. In this paper, we present {\it
XMM-Newton}, Giant Metrewave Radio Telescope (GMRT) and Very Large
Array (VLA) observations and results of a study of the FRII giant
radio source 3C\,457. We describe the radio observations and data
analysis in Section 2, X-ray observations and data analysis in Section
3, observational results in Section 4, discussion in Section 5 and
concluding remarks in Section 6.

\begin{table}
\caption{Radio observing log, which is arranged as follows. Column 1: name of the telescope,
Column 2: the array configuration for the VLA observations, Column 3: frequency of observations,
Column 4: date of observations.}
\begin{tabular}{l c c c c }

\hline
Telescope & Array    & Observed          &Date of Observations       \\
          & Config.  & Frequency         &                           \\
          &          & (MHz)             &                           \\
  (1)     &  (2)     & (3)               & (4)                       \\
\hline
%---------------------------------------------------------------------
GMRT      &        & 239                & 2007 Aug 22 \\
GMRT      &        & 334                & 2007 Sep 01 \\
GMRT      &        & 605                & 2007 Aug 22 \\
VLA$^a$   &  B     & 1477               & 1987 Dec 02 \\
VLA$^a$   &  C     & 4710               & 2000 Jun 12 \\
%---------------------------------------------------------------------
\hline
\end{tabular}  \\
$^a$ archival data from the VLA \\
\label{obslog}
\end{table}

\section{Radio observations and data analysis} 
Both the GMRT and the VLA observations were made in the standard
fashion, with each target source observation interspersed with
observations of the phase calibrator. The primary flux density
calibrator was either 3C\,48 or 3C\,286, with all flux densities being on
the scale of Baars et al. (1977). The total observing time on the
source is a few hours for the GMRT observations while for the VLA
observations the time on source is $\sim$20 minutes. The low-frequency
GMRT data were sometimes significantly affected by radio frequency
interference, and these data were flagged. All the data were analysed
in the standard fashion using the NRAO {\tt AIPS} package. All the
data were self-calibrated to produce the best possible images. The
observaing log for both the GMRT and the VLA observations is given in
Table~\ref{obslog}.

\section{X-ray observations and data analysis} 

3C\,457 was observed with {\it XMM-Newton} in December 2007, with the
medium filter and the pn camera in Extended Full Frame mode for an
observation duration of 53, 53 and 49 ks for the MOS1, MOS2 and pn
cameras, respectively. The data were reduced using the {\it
XMM-Newton} Scientific Analysis Software (SAS) version 7.1, and the
latest calibration files from the {\it XMM-Newton} website. The pn
data were filtered to include only single and double events (PATTERN
$\leq 4$), and \#XMMEA\_EP (excluding bad columns and rows), and the
MOS data were filted according to the standard flag and pattern masks
(PATTERN $\leq 12$ and \#XMMEA\_EM, excluding bad columns and rows).
Filtering for good-time intervals was also applied to exclude regions
affected by background flares, using a lightcurve in the energy range
where the effective area of X-rays is negligible (10 -- 12 keV for
MOS, 12 -- 14 keV for pn); the remaining exposure after GTI-filtering
was 35, 35 and 22 ks for the MOS1, MOS2 and pn cameras, respectively.
Spectra were extracted using the SAS task {\it evselect}, with
appropriate response files generated using {\it rmfgen} and {\it
arfgen}, and spectral analysis was carried out using {\sc xspec}.

\section{Observational results}
\subsection{Radio data}

%%%%%%%%%%%%%%%%%%%%%%%%%%%%%%%%%%%%%%%%%%%%%%%%%%%%%%%%%%%%%%%%%%%%%%%%%
\begin{figure*}
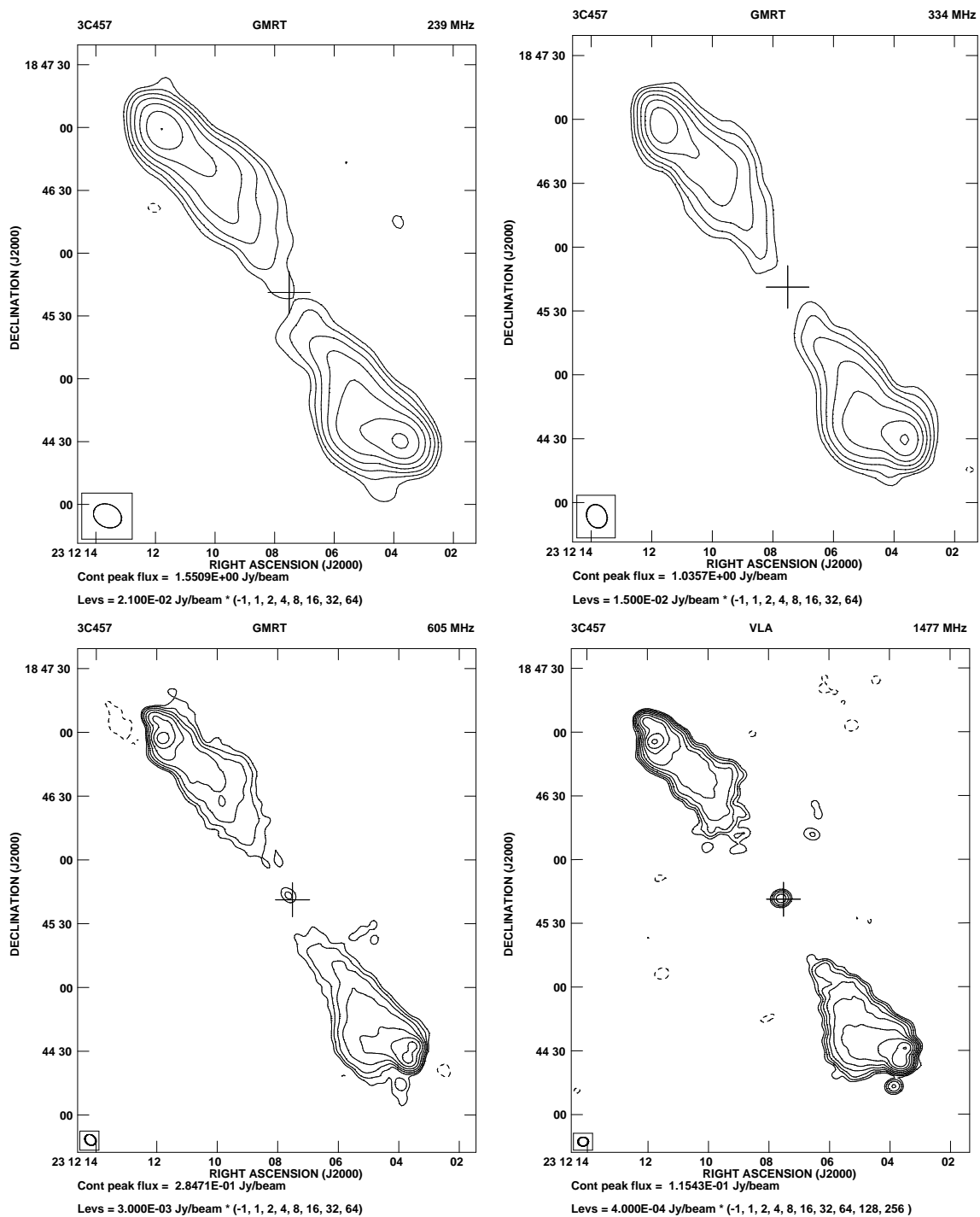

\vbox{
  \hbox{
  \psfig{file=3C457_T.PSC5.PS,width=3in,angle=0}
  \psfig{file=3C457_P.PSC5.PS,width=3.05in,angle=0}
       }
  \hbox{
  \psfig{file=3C457G.IF1P5.PS,width=3in,angle=0}
  \psfig{file=AL146L_2.PS3.PS,width=3in,angle=0}
       }
}

\caption[]{GMRT and VLA images of 3C\,457 at different frequencies,
which are indicated at the top of each image along with the name of
telescope with which it was made. The peak brightness and the contour
levels are given below each image. In all the images the restoring
beam is indicated by an ellipse and the + sign indicates the position
of the optical host galaxy.}
\label{radio_maps}
\end{figure*}
%%%%%%%%%%%%%%%%%%%%%%%%%%%%%%%%%%%%%%%%%%%%%%%%%%%%%%%%%%%%%%%%%%%%%%%%%
The GMRT images of the source at the different frequencies are
presented in Figure \ref{radio_maps}, while the observational
parameters and some of the observed properties are presented in Table
\ref{obs.fluxes}. The radio images of 3C\,457 in Figure~\ref{radio_maps}
show its large-scale structure, with no core detected at 234 and 334
MHz. However, the images at 605, 1477 and 4710 MHz show
a clear detection of a core in addition to the diffuse lobes. The
radio spectra of different components are shown in
Figure~\ref{radio.spect_all.comp}. The core spectral index is
$\sim$0.4 between 605 and 4866 MHz.

The diffuse lobe emission of 3C\,457 clearly shows a backflow
structure and, at low frequencies, fills up the region between the
hotspots and the core. Both the lobes have radio spectra consistent
with a power law within our frequency coverage (see
Figure~\ref{radio.spect_all.comp}). The spectral indices are
1.01$\pm$0.04 and 0.98$\pm$0.04 for the north-eastern lobe (NE lobe)
and south-western lobe (SW lobe) respectively. The spectral index
obtained from the integrated spectrum shown in
Figure~\ref{radio.spect_all.comp} is 0.99$\pm$0.04.

%%%%%%%%%%%%%%%%%%%%%%%%%%%%%%%%%%%%%%%%%%%%%%%%%%%%%%%%%%%%%%%%%%%%%%%%%
\begin{figure}   
\vbox{
  \hbox{
  \psfig{file=AD429C.PS2.PS,width=2.5in,angle=0}
       }
}
\contcaption{}
\end{figure}
%%%%%%%%%%%%%%%%%%%%%%%%%%%%%%%%%%%%%%%%%%%%%%%%%%%%%%%%%%%%%%%%%%%%%%%%%

\begin{figure}
\vbox{
  \hbox{
  \psfig{file=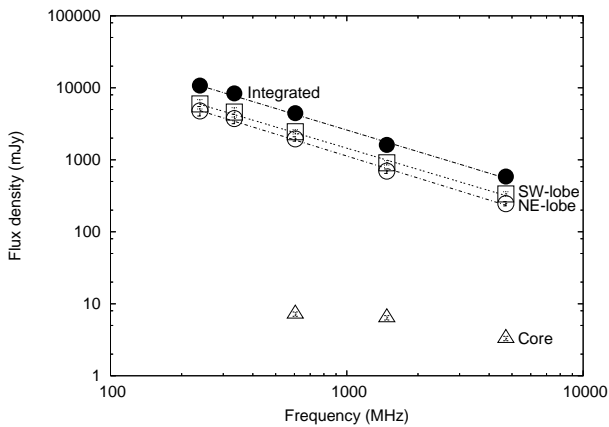,width=3.2in,angle=-90}
       }
}
\caption{ Integrated spectrum and the spectra of all components of
         3C\,457 in radio band. Filled circle: the integrated spectrum.
         Open square: the spectrum of the SW lobe. Open circle: the
         spectrum of the NE lobe. Open triangle: the spectrum of the
         core. In all cases the error bars are smaller than the
         symbols, and are shown within the symbols.}
\label{radio.spect_all.comp}
\end{figure}

\begin{figure}
\vbox{
  \hbox{
  \psfig{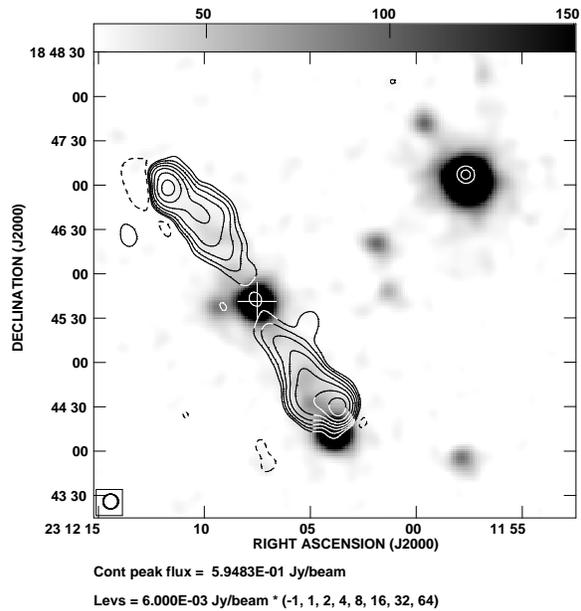}
       }
}
\caption{A 10-arcsec Gaussian smoothed image of 3C\,457 and
its field in the energy range 0.3 to 10 keV made from the combined
MOS1, MOS2 and pn X-ray data is displayed in grey scale. Overlaid on
top of this are the GMRT 10-arcsec 605-MHz radio contours.
The contour levels are displayed at the bottom of the image. The $+$
sign indicates the position of the AGN of the radio source 3C\,457. A
few more AGNs are visible in the field.}
\label{3C457_XnR.overlay}
\end{figure}

\subsection{X-ray data}
\label{xray.data}
Figure~\ref{3C457_XnR.overlay} shows the 605 MHz GMRT contour map of
3C\,457 overlaid on top of the {\it XMM-Newton} image in grey scale in the
energy range 0.3 to 10 keV. The X-ray emission from the AGNs in the
field and the lobes of 3C\,457 are prominent. The three most prominent
AGNs in this field are the core of 3C\,457 (RA: 23 12 07.446$\pm$0.003
Dec: 18 45 40.42$\pm$0.04), an AGN near the southern hotspot of 3C\,457
(AGN-1, RA: 23 12 03.842$\pm$0.003 Dec: 18 44 12.41$\pm$0.05) and an
AGN to the north-west of the host of 3C\,457 at an angular distance
of 140 arcsec (AGN-2, RA: 23 11 57.530$\pm$0.001 Dec: 18 47
03.58$\pm$0.01). All position coordinates are in J2000. All three of
these AGNs have radio counterparts, as evident from
Figure~\ref{radio_maps} and Figure~\ref{3C457_XnR.overlay}.

We extracted spectra from the MOS1, MOS2 and pn files. For each of the
three AGNs mentioned above, we used a small circular extraction region
around it, so that the spectra contain mostly AGN emission with as
little possible contamination from any thermal emission around the AGN
if present. The radii of the spectral extraction circles were
$\sim$16 arcsec, 9 arcsec and 26 arcsec
for the core, AGN-1 and AGN-2, respectively. For the spectral
extraction of the lobes we used rectangular extraction regions
covering the extent of the diffuse radio emission as shown in the GMRT
and VLA images. We also looked for group/cluster scale thermal
emission in the environment around the core of 3C\,457 within a circular
region of radius $\sim$145 arcsec ($\sim$800 kpc). The
circular region for the spectral extraction of the environment should
be large enough to encompass most of the environment around the radio
galaxy, and so would typically encircle the lobes and extend somewhat
beyond. However, since 3C\,457 is a giant radio galaxy, a circular
region enclosing the entire source is in this case more than enough as
most of the environmental emission comes from the central region. It
is not important whether we choose a region of 145 arcsec or
much larger, as the environment follows a $\beta$-model profile and
falls off rapidly beyond $\sim 20$ arcsec (see
Figure~\ref{beta-model.plot}). Spectral extraction regions are all
shown in Figure~\ref{spect.extr.reg}. For spectral analysis of the
AGNs, local background subtraction was used. Details of background
subtraction for the lobe regions and the hot gaseous environment is
discussed later. A scaling has been done for the source and background
counts to account for the difference in areas between the source
extraction and background subtraction regions.

The value of the Galactic neutral hydrogen column density
($N_{\rm H}=0.05\times 10^{22}$ cm$^{-2}$), as obtained from Dickey \&
Lockman (1990), does not yield overall acceptable fits to the spectra
of all of the X-ray sources in the field of view with plausible
models. The Galactic extinction measurement from Schlegel, Finkbeiner
\& Davis (1998) as well as our data suggest a higher value of $N_{\rm H}$
along the line of sight (LOS) towards 3C\,457. Using the formula
$N_{\rm H}= 5.9
\times 10^{21} \times E_{B-V}$ cm$^{-2}$ (Spitzer 1978) and
substituting $E_{B-V}$ (colour excess in mag) of 0.258 mag, as obtained
from Schlegel et al. (1998),
%NASA Extragalactic Database (NED) 
we get $N_{\rm H}= 0.15\times 10^{22}$ cm$^{-2}$, which is higher than
the value obtained from Dickey \& Lockman (1990). As a further
investigation of the appropriate value for the $N_{\rm H}$ we also
modelled the spectra of AGN-1 and AGN-2, which are within 165 arcsec
of the core, by fitting a single power law (wabs(pow) in {\sc xspec}).
The fits are acceptable and we get very similar best-fitting values of
$N_{\rm H}$, which are 0.238$^{+0.026}_{-0.026}\times 10^{22}$ and
0.225$^{+0.013}_{-0.011}\times 10^{22}$ cm$^{-2}$ for AGN-1 and AGN-2
respectively. These values of $N_{\rm H}$ will account for the
intrinsic as well as Galactic $N_{\rm H}$. Therefore, we consider the
maximum value of Galactic $N_{\rm H}$, neglecting the error in our
model fits, to be 0.238$\times 10^{22}$ cm$^{-2}$. It is unlikely that
both AGN-1 and AGN-2 will conspire to have similar intrinsic column
density to yield the similar (within errors) total column density, and
so it is quite plausible that the intrinsic column density of AGN-1
and AGN-2 are small compared to the Galactic $N_{\rm H}$. The fits are
acceptable for values of $N_{\rm H}$ between 0.15$\times 10^{22}$ and
0.238$\times 10^{22}$ cm$^{-2}$. As AGN-2 is stronger and the fit is
better, we adopted an $N_{\rm H}$ value of 0.225$\times 10^{22}$
cm$^{-2}$ and freeze it for all the spectral fits described below.
However, we discuss later in the paper how our results are modified if
the $N_{\rm H}$ is instead 0.15$\times 10^{22}$ cm$^{-2}$, which
yields best-fit parameters for the X-ray AGN that are not completely
unacceptable/unphysical.

\subsubsection{Spectrum of the core}
Our initial fitting of a single power law to the core spectrum was
unacceptable ($\chi_{red}^2$= 3.88), and so multiple components are
obviously required. We therefore fitted the spectrum with two
composite models: Model I -- a double power law (a soft power law with
Galactic absorption plus a hard absorbed power law;
wabs(pow+zwabs(pow)) in xspec), and Model II -- a soft mekal model
plus a hard absorbed power law (wabs(mekal+zwabs(pow)) in xspec).
Since 3C\,457 is a radio galaxy it might be expected to have a double
power-law spectrum, with one component related to the accretion
process in the nucleus and the other related to the emission from the
jet base (e.g. Hardcastle, Evans \& Croston 2006), which is the physical
motivation for Model I. We also know that there is a hot gaseous
medium around any radio galaxy (see Miller et al. 1999; Hardcastle \&
Worrall 1999; Worrall \& Birkinshaw 2000; Croston et al. 2003; Belsole
et al. 2004; Belsole et al. 2007) which emits via thermal
bremsstrahlung. Model II allows us to investigate whether this thermal
component is prominent relative to the nuclear component. Both the
model fits are equally good. The best-fitting parameters for both the
models are given in Table~\ref{fitstat_core} which is self
explanatory.

\subsubsection{Spectra of the lobes}
We then extracted the spectrum of the SW lobe. The rectangular
spectral extraction region of the SW lobe is shown in Figure
\ref{spect.extr.reg}. We choose the rectangle to minimise any possible
contribution from the core and AGN-1. The background region was placed
at the same distance from the core so as to minimise the error in
background subtraction. We separately fitted mekal and power law
models to the spectrum of the SW lobe. Both the fits have similar
$\chi_{red}^2$ value. However, the best-fitting temperature of the mekal
model is high, which is consistent with the result obtained by Croston
et al. (2004) for the radio source 3C\,223. The single power law and
mekal model fit parameters are in Table \ref{fitstat_slobe.n.env}. For
the north-eastern radio lobe (NE lobe) it is not possible to carry out
spectral fits as the number of counts is small, partly because of
overlap with the chip gaps of the pn camera. However, we determined
the 1-keV flux density of the NE lobe so as to enable a magnetic field
calculation (see Sec \ref{Lobe-emission}). We used the ratio of total
MOS1 and MOS2 counts for the NE lobe to SW lobe ratio to scale the
1-keV flux density of SW lobe obtained from the fit to determine the
1-keV flux of NE lobe.

\subsubsection{Spectrum of the hot gaseous environment}
\begin{figure*} 
%\begin{figure}
  \hbox{
  \psfig{file=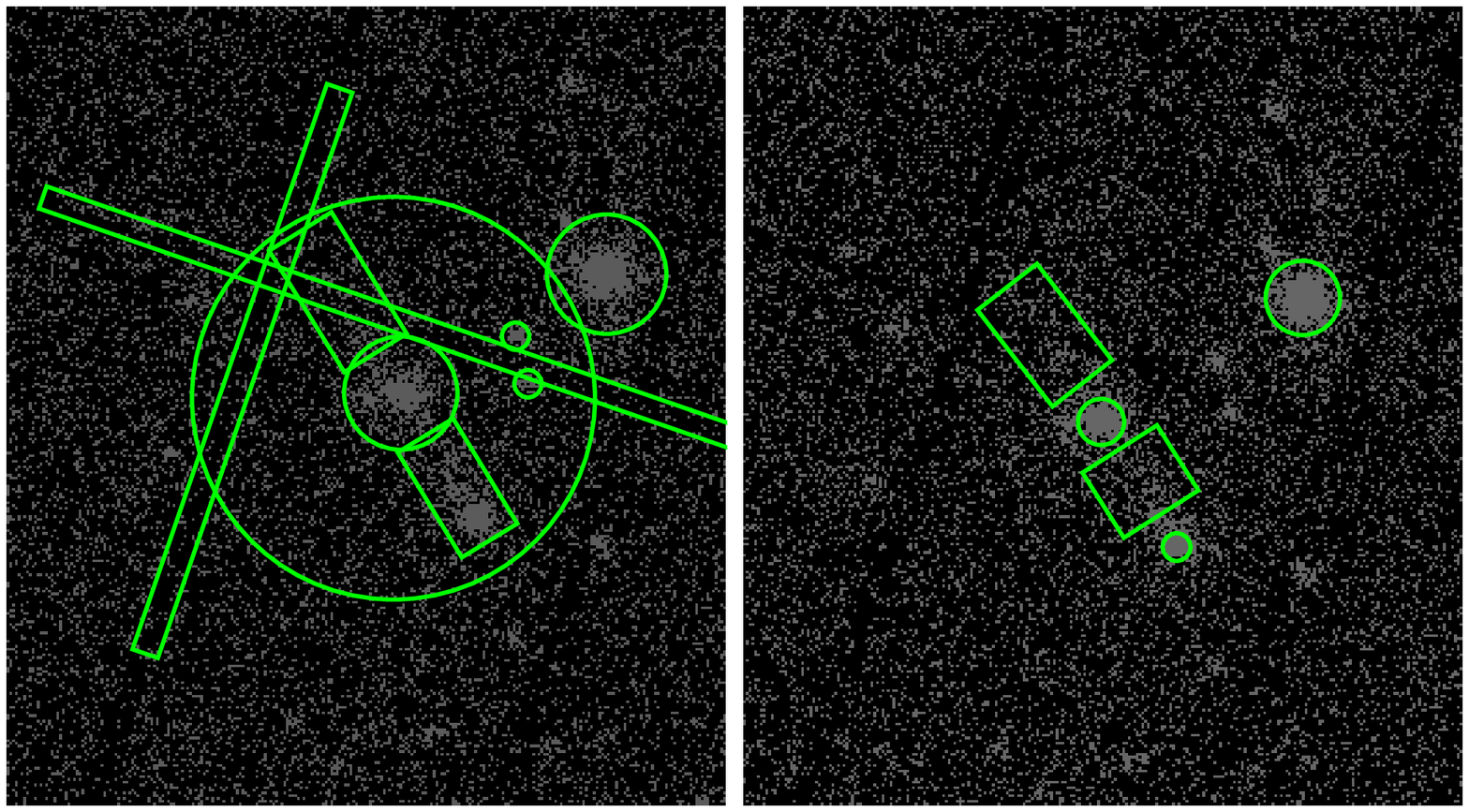,width=5.0in,angle=0}
       }
\caption{Left panel: Spectral extraction region (largest circle) for
         the environment around 3C\,457. The other circles and the
         rectangles are to subtract the emission from the core of
         3C\,457, unrelated AGNs, lobes and areas affected by the chip
         gaps. Right panel: Spectral extraction region for the lobes
         (rectangles), core (circle in between the lobes) and the
         unrelated AGNs (other circles). }
\label{spect.extr.reg}
%\end{figure}
\end{figure*}

We examined any possible thermal gas environment around the core of
3C\,457. We extracted the spectrum within a large circle of radius of 
about 145 arcsec ($\sim$ 800 kpc) with the position of the core as
the centre. We excluded the core and two lobes of 3C\,457, and all the
AGNs that are within the extraction region (see Figure
\ref{spect.extr.reg}) for the environment. We have also excluded any
AGNs that are outside but close to the extraction region. For
background subtraction we used a circular background region of radius
$\sim$90 arcsec at a distance of $\sim$350 arcsec
from the core of 3C\,457 towards a position angle of
$\sim$165$^{\circ}$. The background region is sufficiently far from
the core of 3C\,457 and any other AGN so that it is unlikely to contain
any emission from hot gaseous medium or other AGNs. We fitted the
spectrum of only the environment with the mekal model at a redshift
($z$) of 0.428 with the frozen values of $N_{\rm H}=0.225 \times 10^{22}$
cm$^{-2}$ and metal abundance = 0.35 $Z_{\odot}$. This yields a
reasonable fit with the gas temperature equal to
2.62$^{+1.15}_{-0.69}$ keV. The best-fitting parameters are shown in Table
\ref{fitstat_slobe.n.env}.

%=======================================================================
%%%%%%%%%%%%%%%%%%%%%%%%%%%%%%%%%%%%%%%%%%%%%%%%%%%%%%%%%%%%%%%%%%%%%%%%%

%%%%%%%%%%%%%%%%%%%%%%%%%%%%%%%%%%%%%%%%%%%%%%%%%%%%%%%%%%%%%%%%%%%%%%%%%
\begin{table*}
\caption{ The observational parameters and observed properties of the
sources are presented in this table, which is arranged as follows.
Column 1: Name of the source; column 2: frequency of observations in
units of MHz, with the letter G or V representing either GMRT or VLA
observations; columns 3-5: the major and minor axes of the restoring
beam in arcsec and its position angle (PA) in degrees; column 6: the
rms noise in units of mJy beam$^{-1}$; column 7: the integrated flux
density of the source in mJy estimated by specifying an area enclosing
the entire source. We examined the change in flux density by
specifying different areas and found the difference to be within a few
per cent. The flux densities at different frequencies have been
estimated over similar areas. Columns 8, 11 and 14: component
designation, where NE, SW and C denote north-eastern, southern-western
and core components respectively; columns 9 and 10, 12 and 13, and 15
and 16: the peak and total flux densities of each of the components in
units of mJy beam$^{-1}$ and mJy respectively.}

\begin{tabular}{l l rrr r r r rr l rr r rr}
\hline
Source   & Freq.       & \multicolumn{3}{c}{Beam size}                    & rms      & S$_I$   & Cp  & S$_p$  & S$_t$  & Cp   & S$_p$ & S$_t$ & Cp  & S$_p$   & S$_t$     \\

           & MHz         & arcsec & arcsec & $^\circ$ &    mJy   & mJy     &     & mJy    & mJy    &      & mJy   & mJy   &     & mJy     & mJy       \\
           &             &                   &                   &          & beam$^{-1}$&       &     &beam$^{-1}$   &        &      & beam$^{-1}$    &       &     & beam$^{-1}$      &           \\ 
   (1)     & (2)   & (3)  & (4)  & (5)  & (6)  & (7)  &(8)& (9)  & (10) & (11)  &   (12)  &(13)  &(14)& (15) & (16)  \\
\hline

3C\,457      & G239  & 13.89 & 10.83 & 64   & 4.22 & 10757 & NE & 1350 & 4772 &       &           &       & SW & 1551 & 5976   \\
           & G334  & 11.29 &  9.32 & 24   & 3.02 &  8358 & NE &  943 & 3740 &  C    &      $<$12  &       & SW & 1036 & 4612   \\
           & G605  &  5.60 &  4.57 & 45.5 & 0.61 &  4442 & NE &  285 & 1953 & C$^g*$& 7.2       & 7.2   & SW &  266 & 2451   \\
           &V1477  &  4.82 &  4.17 & 273  & 0.08 &  1610 & NE &  115 &  696 & C$^g$ & 5.9       & 6.4   & SW &  107 &  909   \\
           &V4710  &  4.91 &  4.56 &  56  & 0.04 &   585 & NE &   41 &  246 & C$^g$ & 3.2       & 3.3   & SW &   44 &  336   \\ 
       
\hline
\end{tabular} 
$^g$ Flux density has been determined from the two dimensional Gaussian fit. \\
$*$  Flux density has been determined by re-mapping with lower uv-cutoff to remove the contamination due to diffuse emission. \\
\label{obs.fluxes}
\end{table*}
%%%%%%%%%%%%%%%%%%%%%%%%%%%%%%%%%%%%%%%%%%%%%%%%%%%%%%%%%%%%%%%%%%%%%%%%%

%%%%%%%%%%%%%%%%%%%%%%%%%%%%%%%%%%%%%%%%%%%%%%%%%%%%%%%%%%%%%%%%%%%%%%%%%
     
\begin{table*}
\caption{Fit statistics of the spectrum of the core}
\begin{tabular}{llllccc}
\hline\hline
Model          & Parameter                 & Model I                        &    Model II           \\
component      &                           &                                &                       \\
  (1)          &  (2)                      & (3)                            &     (4)                \\
\hline
%------------------------------------------------------------------------------------------------------------
Soft power law & $\Gamma$                  & 2.18$^{+0.32}_{-0.29}$         &                        \\
               & 1 keV flux density        & 4.82$^{+0.42}_{-0.42}$         &                        \\
               & (nJy)                     &                                &                        \\\hline
%------------------------------------------------------------------------------------------------------------
Soft mekal     & kT (keV)                  &                                & 2.11$^{+2.00}_{-0.54}$  \\
               & Unabsorbed flux           &                                &(2.56$^{+20.87}_{-0.25}$)$\times10^{-14}$ \\
               & (erg/s/cm$^2$)            &                                &                      \\ \hline
%------------------------------------------------------------------------------------------------------------
Hard power law & Nuclear $N_{\rm H}$ (cm$^{-2}$) &(30.0$^{+6}_{-6}$)$\times10^{22}$ & (26.0$^{+9.5}_{-4.5}$)$\times10^{22}$ \\
               & $\Gamma$                  & 1.58$^{+0.19}_{-0.13}$           & 1.45$^{+0.11}_{-0.11}$                \\
               & Unabsorbed flux  &(9.72$^{+2.22}_{-8.68}$)$\times10^{-13}$ &(9.31$^{+1.97}_{-7.75}$)$\times10^{-13}$   \\
               & (erg/s/cm$^2$)            &                                        &                       \\ \hline
$\chi_{red}^2$ &                           & 0.69                                   &  0.72                 \\
%------------------------------------------------------------------------------------------------------------
\hline
\end{tabular} \\
Note: The observed part of the spectrum has been integrated to find the flux.
\label{fitstat_core}
\end{table*}

\begin{table*}
\caption{Fit statistics of the spectrum of SW lobe and environment}
\begin{tabular}{llllccc}
\hline\hline
 Source              & Model          & Parameter              & Best-fitting values    \\
 Component           &                &                        &                     \\
  (1)                &  (2)           & (3)                    &  (4)               \\
\hline
%------------------------------------------------------------------------------------------------
  SW lobe             & Power law      & $\Gamma$               & 1.63$^{+0.23}_{-0.22}$   \\
                     &                & 1 keV flux density     & 3.91$^{+0.50}_{-0.51}$   \\
                     &                &   (nJy)                &                          \\
                     &                & $\chi_{red}^2$         & 0.85                     \\
%------------------------------------------------------------------------------------------------
                     & mekal          & kT (keV)               & 8.29$^{+14.54}_{-3.59}$  \\
                     &                & Unabsorbed flux        &(3.60$^{+0.69}_{-1.13}$)$\times 10^{-14}$ \\
                     &                & (erg/s/cm$^2$)         &                          \\
                     &                &$\chi_{red}^2$          & 0.95                     \\ \hline
%------------------------------------------------------------------------------------------------
 Environment         & mekal          & kT (keV)               & 2.62$^{+1.15}_{-0.69}$        \\
                     &                & Unabsorbed flux        &(5.98$^{+1.14}_{-1.25}$)$\times 10^{-14}$  \\
                     &                & (erg/s/cm$^2$)         &                             \\
                     &                &$\chi_{red}^2$          & 1.29                        \\
\hline
\end{tabular} \\
Note: The observed part of the spectrum has been integrated to find the flux.
\label{fitstat_slobe.n.env}
\end{table*}
%%%%%%%%%%%%%%%%%%%%%%%%%%%%%%%%%%%%%%%%%%%%%%%%%%%%%%%%%%%%%%%%%%%%%%%%%

\section{Discussion and results}

\subsection{X-ray emission from different components}

\subsubsection{Core emission}
%{\Large \bf Core emission:}
The radio core of 3C\,457 is a flat spectrum core as is evident from
Figure~\ref{radio.spect_all.comp}. The X-ray spectral analysis of
3C\,457 reveals that the core spectrum is well fitted with two two-component
models. If we adopt Model I, then the soft power law component of the
core spectrum originates from the base of the jets of the
radio galaxy 3C\,457, which is consistent with the results of Croston et
al. (2004). The 5-GHz core flux of this source from our measurement is
3.30$\pm$0.16 mJy (a 5 per cent error in flux for VLA data is assumed)
and the 1-keV X-ray flux obtained from the fit to Model I corresponds
to 4.82$\pm$0.42 nJy. So it is consistent with the radio and X-ray
flux correlation of the cores of radio galaxies, as found by
(e.g.) Hardcastle \& Worrall (1999). If we adopt Model II, then the best-fitting
temperature (2.11$^{+2.00}_{-0.54}$ keV) of the soft thermal component
is similar to that of the hot gas environment of 3C\,457. Its bolometric
luminosity (0.1$-$100 keV) is $\sim$2.04$^{+67.2}_{-0.2}$ $\times$
10$^{43}$ erg s$^{-1}$. This mekal temperature and the bolometric
luminosity is consistent with the $L_{\rm X}$--$T_{\rm X}$ correlation for the
cluster atmosphere (Osmond \& Ponman 2004). However, the bolometric
luminosity is not well constrained and also this luminosity is only
from a fraction of the environment around the core.

In Model I the soft power law is well constrained and consistent with
earlier studies (e.g. Croston et al. 2004); whereas, in Model II the
soft mekal model is not well constrained. Therefore, we favour Model I
for the core spectrum, which is consistent with the results found for
the nuclear spectra of 3C\,223 and 3C\,284 (Croston et al. 2004 and
references therein). There is no evidence of a redshifted 6.4 keV iron
line in our core spectrum.

\begin{figure}
\vbox{
  \hbox{
  \psfig{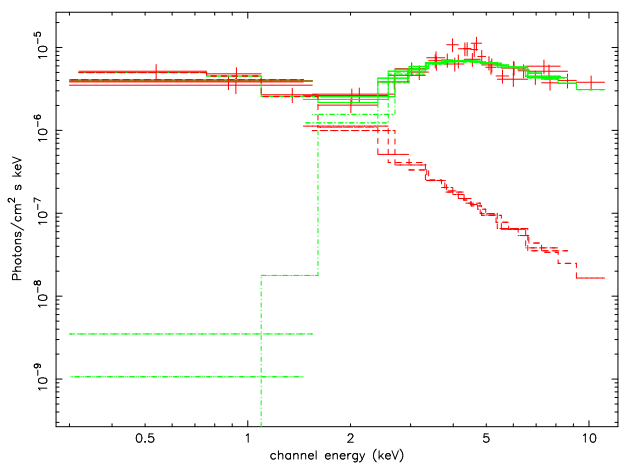}
       }
}
\caption{X-ray spectrum of the core of 3C\,457. The spectrum was fitted with a
soft power law plus a hard absorbed power law (wabs(pow+zwabs(pow)) in
{\sc xspec}. This composite model is fitted separately with the data
from the three cameras (MOS1, MOS2 and pn) and plotted in three dashed
lines.}
\label{specage.vs.dist.plot}
\end{figure}

\subsubsection{Lobe emission}
\label{Lobe-emission}
We fitted single power law (wabs(pow) in {\sc xspec}) and mekal models
(wabs(mekal) in {\sc xspec}) to the X-ray spectrum of the SW lobe.
Both the fits are equally good, with $\chi_{red}^2$ equal to 0.85 and
0.95 for the power law and mekal model respectively. But the mekal
model temperature is too high and very poorly constrained. One
plausible reason for such a high temperature, as pointed out by
Croston et al. (2004), is due to supersonic expansion of the
lobes which shock-heats their immediate environment. However, the
single power law model yields a better constrained photon index
($\Gamma=$1.63$^{+0.23}_{-0.22}$). If we assume that the X-rays from
the lobes are due to IC-CMB (as it is a power law), then this photon
index is consistent with $\alpha_{inj}=$0.82, as constrained from the
radio measurements (Jamrozy et al. 2008). Moreover, the morphology is
consistent with IC-CMB model, i.e., in the lobes of X-ray image, there
is no edge brightened structure, which might be expected for shock
heating by the supersonically expanding lobes. So we adopt the
power-law interpretation of the spectrum, although we discuss the
thermal model for the X-ray emission from the lobes in
Section~\ref{pressure-balance}. We have constrained the magnetic
fields of the lobes, using the {\sc synch} code (Hardcastle,
Birkinshaw \& Worrall 1998; Hardcastle et al. 2004), 
from IC-CMB modelling without the assumption of
equipartition/minimum-energy. The X-ray emission due to the IC-CMB
process in the 0.1--10 keV range is due to scattering by electrons of
Lorentz factor ($\gamma$) around 1000. For the typical magnetic fields
of radio lobes, those electrons radiate at a low radio frequency, i.e.
a few tens to a few hundreds of MHz. Therefore, for our IC-CMB
modelling we need to either assume a fiducial value for the
low-frequency spectral index (the injection spectral index
($\alpha_{inj}$)) of each lobe or to constrain it. We have taken
$\alpha_{inj}$=0.82 for both the lobes, as constrained by Jamrozy et
al. (2008). The magnetic fields are 0.68$^{+0.06}_{-0.09}$ and
0.40$^{+0.04}_{-0.02}$ nT for the NE and SW lobes respectively. The
corresponding minimum energy fields are 0.86 and 0.82 nT, which are
within a factor of two only, in good agreement with the results of
Croston et al. (2005) for a large sample of radio galaxies. Minimum
energy fields are calculated with the assumptions that the filling
factor of lobes are unity, the energetically dominant particles are
the radiating particles only (the contribution of protons has been
neglected) and the electron energy spectra extend from $\gamma=$10 to 10$^{5}$.

\begin{figure} 
\vbox{
  \hbox{ 
  \psfig{file=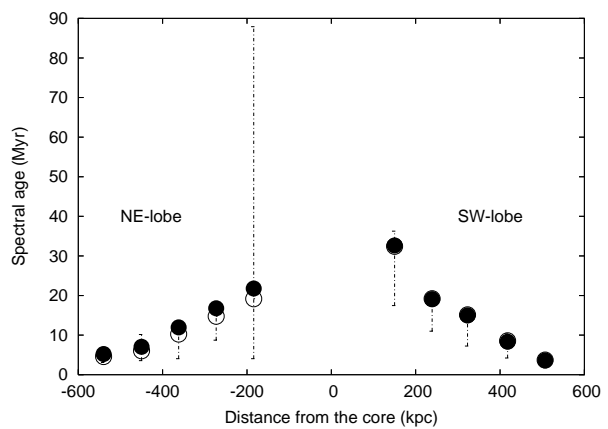,width=3.2in,angle=-90}
       }
}
\caption{Spectral age vs. distance plot. Filled circle: spectral age,
         using classical equipartition magnetic field, as in Jamrozy
         et al. (2008). Open circle: spectral age wih error bars,
         using the field determined from IC-CMB modelling.}
\label{specage-dist.plot}
\end{figure}

Spectral ageing analysis, using the revised equipartition magnetic
field (see Beck \& Krause 2005), done by Jamrozy et al. (2008) has
revealed a prominent curvature in the spectral age vs. distance plot.
Whereas, the same plot with the classical minimum energy field does
not show such prominant curvature (see both Figure~9 of Jamrozy et al.
2008 and Figure~\ref{specage-dist.plot} of this paper). The revised
equipartition magnetic field ($B_{\rm eq}$(rev)) is obtained from the
formalism of Beck \& Krause (2005). Their formula (equation A18) has
the parameter {\bf K}$_{0}$ which is the ratio of the number density
of protons to that of electrons in the energy range where losses are
small. It is relevant to note that in this formalism particle energy
is dominated by the protons. Estimating {\bf K}$_{0}$ from the
equation {\bf K}$_{0}$=$(\frac{m_{\rm p}}{m_{\rm e}})^{\alpha}$ as given by their
equation (7) (where $m_{\rm p}$ is the proton mass, $m_{\rm e}$ is the
electron mass and $\alpha\approx \alpha_{\rm inj}$ which depends on
the low-frequency spectral index in the observed synchrotron
spectrum), we can constrain the proton spectrum and hence estimate the
revised equipartition magnetic field strength, $B_{\rm eq}$(rev). They
have used the equipartition magnetic field using both the classical
and revised formalisms (see Konar et al. 2008) for each strip to
estimate the spectral age. It is worth mentioning that in the usual
classical equipartition formalism (as used by Jamrozy et al. 2008) the
electrons and protons have same total kinetic energy, which is a mere
assumption and has no observational evidence. However, in this paper
we have calculated our classical equipartition magnetic field for the
source 3C\,457 with the assumption that the energetically dominant
particles in the lobes are the radiating particles only (Croston et
al. 2005). We have estimated the spectral age of each strip of the
lobes, using our magnetic field derived from IC-CMB modeling of the
lobe X-ray emission (described in the beginning of this section), and
compared with the values obtained from Jamrozy et al. (2008) for the
classical equipartition field. We plotted the spectral ages estimated
by us as well as Jamrozy et al. (2008) with the distance of the strip
from the core. Unlike Jamrozy et al. (2008), we have used the constant
field for all the strips, because IC-CMB modelling is not possible for
each strip separately. Our estimates of ages for all the strips with
constant magnetic field determined from IC-CMB modelling are close to
the values obtained by Jamrozy et al. (see
Figure~\ref{specage-dist.plot}). In fact for the SW lobe our points
lie exactly on top of theirs. This shows that the classical
equipartition magnetic field yields reasonably accurate spectral age,
provided the break frequency determination is accurate.

All the calculations regarding magnetic field and pressure have been
done with the assumption that the lobes are in the plane of the sky,
as 3C\,457 is not a quasar. However, assuming that the source is at
45$^{\circ}$ with the LOS, our calculations yield very similar
numbers. The minimum energy fields change by less than $\sim$10 per cent from
the values quoted above, and the fields constrained from the IC-CMB
modelling changes by much less than a few per cent and hence much less
than the error bars quoted above. Our estimatation of spectral age
also yields very similar results. Hence all our results about magnetic
fields and spectral ages are similar even if 3C\,457 is inclined at an
angle of 45$^{\circ}$ to the LOS.

\begin{figure}
\vbox{
  \hbox{
  \psfig{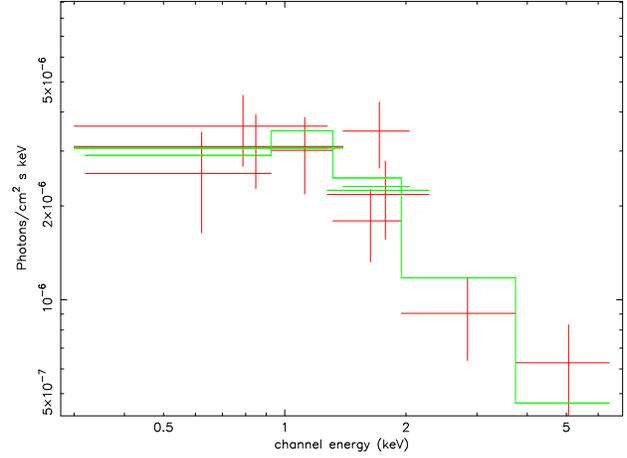}
       } 
}
\caption{X-ray spectrum of SW lobe of 3C\,457, fitted with a power-law model.}
\label{specage.vs.dist.plot}
\end{figure}

\begin{figure}
\vbox{
  \hbox{
  \psfig{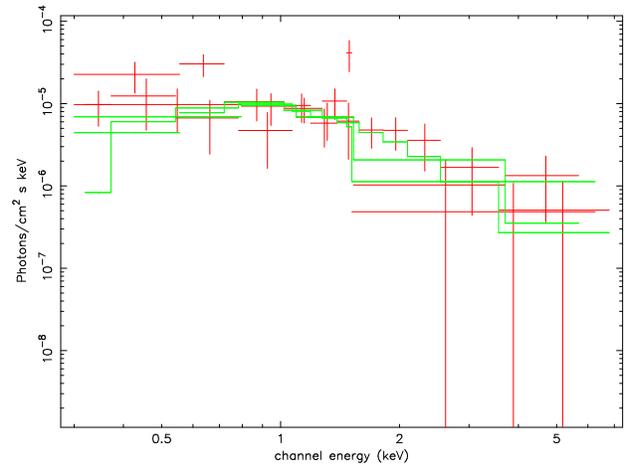}
       }
}
\caption{X-ray spectrum of environment of 3C\,457. This is fitted with
         the mekal model. Three different lines indicate that the data
         from each of the three camera (MOS1, MOS2 and pn) are fitted
         separately.}
\label{specage.vs.dist.plot}
\end{figure}

\subsubsection{Emission from the environment}
We have made an unambiguous detection of a poor-cluster scale hot
gaseous environment around the core of the radio galaxy 3C\,457 over a
circular region of radius $\sim$800 kpc. The environment is consistent
with an isothermal gas. The spectrum is a good fit to the mekal model.
The best-fitting temperature is 2.62$^{+1.15}_{-0.69}$ keV and so is
reasonably well constrained. The unabsorbed bolometric luminosity
(within the 0.1-100 keV energy band) of the environment, excluding the
regions of lobes and a circular region of radius 40 arcsec
around the core, is $\sim$(4.65$^{+1.21}_{-1.44}$)$\times$10$^{43}$
erg s$^{-1}$. This needs to be corrected by a factor taking the
excluded regions into account. We determined the radial surface
brightness profile for the environment around the radio source. The
radial surface brightness in units of counts arcsec$^{-2}$ was
extracted from concentric annuli with point sources, chip gaps and
lobes masked out. We fitted a model consisting of a point source
situated at the position of the core of 3C\,457 and a single $\beta$
model. The Bayesian estimates of $\beta$ and core radius ($r_c$) are
$\beta = 0.51^{+0.02}_{-0.17}$ and $r_c = 11.2_{-6.4}^{+22.7}$. Errors
are the 1-$\sigma$ credible interval. The detailed analysis of the
spatially extended emission around the core of 3C\,457 and the
$\beta$-model fitting procedure were carried out in the same way as
described in Croston et al. (2008).
%%%%%%%%%%%%%%%%%%%%%%%%%%%%%%%%%%%%%%%%%%%%%%%%%%%%%%%%%%%%%%%%%%%%%%%%%%%%%%%%%%%%%%
\begin{figure}
\vbox{
  \hbox{
  \psfig{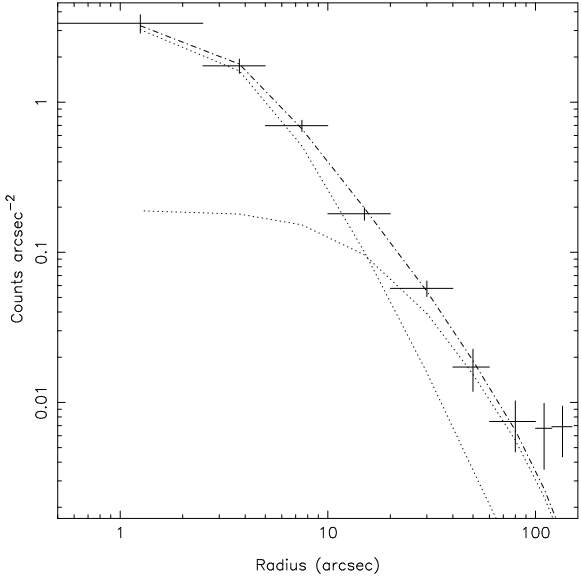}
       }
}
\caption{The surface brightness profile of the environment of 3C\,457.
The dot-dashed line is the total model (a point source + a
$\beta$-model) fitted to the data from all three cameras and dotted
lines are two components of the total model. The crosses are the pn
data only. It appears to be a poor fit in the outer couple of bins
since the data from three cameras disagree about the significance of
any emission on this scale.}
\label{beta-model.plot}
\end{figure}
%%%%%%%%%%%%%%%%%%%%%%%%%%%%%%%%%%%%%%%%%%%%%%%%%%%%%%%%%%%%%%%%%%%%%%%%%%%%%%%%%%%%%%
By extrapolating the $\beta$ model back to the core of 3C\,457, and
taking into account the masking of the lobe and core regions we have
estimated and applied the correction factor to the bolometric X-ray
luminosity of the environment given above. The corrected bolometric
luminosity is $\sim$1.51$^{+1.33}_{-0.97}\times$10$^{44}$ erg
s$^{-1}$. The environment of 3C\,457 seems to follow the luminosity $-$
temperature correlation (hereafter $L_{\rm X}$-$T_{\rm X}$ correlation, Osmond
\& Ponman 2004). Their best-fitting relationship between $L_{\rm X}$ and
$T_{\rm X}$ is
\begin{equation}
\log L_X = (3.26\pm0.12)\log T_X + (42.44\pm0.10).
\label{Lx-Tx_corrln}
\end{equation}
The temperature of the system predicted from
Equation~\ref{Lx-Tx_corrln} is 3.42$^{+1.29}_{-1.17}$ keV. Our best
fit temperature of this cluster scale environment is
2.62$^{+1.15}_{-0.69}$ keV which is similar to what is predicted from
the $L_{\rm X}$-$T_{\rm X}$ correlation. Given the error bars of the
temperatures we cannot say whether the expansion of lobes has any
large-scale heating effect due to $P{\rm d}V$ work on the environment.
Moreover, it should be noted that the best-fiting value of our measured
temperature of the environment is lower than the mean value of what is
predicted from the $L_{\rm X}$-$T_{\rm X}$ correlation, i.e.,
Equation~\ref{Lx-Tx_corrln}. Therefore, it is unlikely that there is
any significant global heating effect on the environment due to $P{\rm d}V$
work by the lobes. The number of counts are not sufficient to check
whether there are any temperature variations in the atmosphere.

\subsection{Pressure balance between radio lobes and environments}
\label{pressure-balance}
\begin{figure}
\vbox{
  \hbox{
  \psfig{file=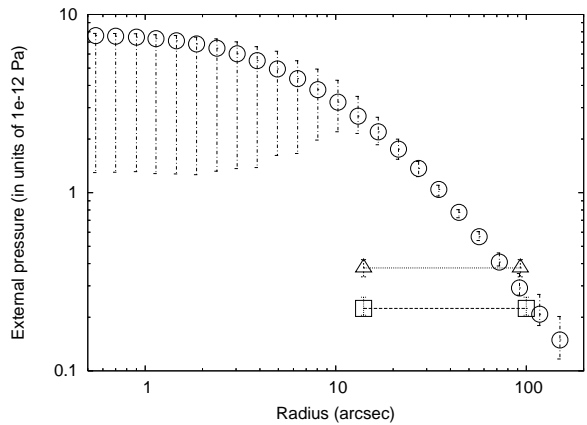,width=3.2in,angle=-90}
       }
}
\caption{Pressure vs. deprojected radius plot of the environment of
         3C\,457 at the best-fitting temperature of 2.62 keV. The
         deprojected radius (on the x-axis) is measured from the
         position of the core of 3C\,457. Open circles: pressure of the
         environment vs. radial distance from the core. Open squares:
         pressure of the NE lobe vs. radial distance from the core.
         Open triangles: pressure of the SW lobe vs. radial distance
         from the core. The lobes are assumed to be in the plane of
         the sky in this plot. The error in pressure shown in this
         plot is only due to the error in measurement of the emission
         measure only. Errors on temperature are not shown for clarity.}
\label{press.vs.radius}
\end{figure}

Radio lobes are thought to be filled with mostly relativistic plasma,
and are generally found to be close to the equipartition condition, if
we assume that the lobes contain mostly radiating particles (Croston
et al. 2005; Kataoka \& Stawarz 2005). Using the {\sc synch} code
(Hardcastle et al. 1998, 2004), we have determined the pressure of the
entire lobe through IC-CMB modelling. For the internal pressure
calculation, we have assumed that the filling factor of the lobes is
unity, the energetically dominant particles are the radiating
particles only (the contribution of protons has been neglected) and
the radio spectra extend from $\gamma=$10 to 10$^{5}$ with
$\alpha_{inj}$=0.82 as constrained by Jamrozy et al. (2008). The
pressures are $0.224^{+0.034}_{-0.019} \times 10^{-12}$ and
$0.378^{+0.043}_{-0.042}\times 10^{-12}$ Pa for the NE and SW lobe
respectively. As the lobes contain mostly ultra-relativistic plasma,
the sound speed will be very high ($=\frac{c}{\sqrt{3}}$). As a result
we can assume that the entire lobe has the same pressure. We have
determined the pressure of the environment as a function of radius
from the core of 3C\,457 from the emission measure and $\beta$-model fit
parameters. The pressure vs. radius plot is shown in
Figure~\ref{press.vs.radius}. The line connecting two triangles show
the pressure along the entire SW lobe detected in low frequency GMRT
maps. The line joining two squares show the same for the NE lobe. From
the figure it is clear that the lobes are approximately at pressure
balance with the environments near the hotspots. At the lobe-head the
ratios of internal to external pressures are $\sim$
0.84$^{+0.43}_{-0.25}$ and 1.31$^{+0.65}_{-0.40}$ for the NE and
SW lobes respectively. At the mid points of the lobes, the ratio of
external to internal pressures are $\sim$2.51$^{+1.18}_{-0.70}$ and
$\sim$1.64$^{+0.75}_{-0.47}$ for the NE and SW lobes respectively. Given
the error bars, the mid points of both the lobes are clearly
under-pressured, although the mid point of the SW lobe might be close
to pressure balance with the environment.

The presence of the hotspots says that the lobes are still expanding
very fast in the longitudinal direction. There is likely to be ram
pressure balance at the lobe-heads. We can calculate the head velocity
of the lobes from ram pressure balance. The ram pressure balance
equation is
\begin{equation}
 n_{\rm p}\mu m_{\rm p} v_{\rm head}^2 = \frac{L_{\rm jet}}{v_{\rm jet}A_{\rm hs}}
\label{rampressure.eqn_1}
\end{equation}
where $n_{\rm p}$ is the particle density of the environment, $\mu$ is the mean
molecular mass of particles in the environment, $m_{\rm p}$ is the proton
mass, $v_{\rm head}$ is the jet head velocity, $L_{\rm jet}$ is the jet power
and can be expressed as $\frac{U_{\rm tot}}{t}$ with $t$=spectral age,
$v_{\rm jet}$ is the jet bulk speed which is usually assumed to be
close to the speed of light and $A_{\rm hs}$ is the hotspot area over
which the jet is impinging the environment. Putting
$v_{jet}=c=3\times10^{10}$ cm~s$^{-1}$, $\mu=$1.4 and expressing all
other quantities in the practical units, the working formula for
$v_{head}$ in units of c ($\beta_{\rm head}$) is given by
\begin{equation}
\beta_{\rm head}^2= 5.31\times 10^{-5} \frac{(\frac{U_{\rm
      tot}}{10^{60}erg}) }{(\frac{t}{Myr})(\frac{n_{\rm p}}{cm^{-3}})
                          (\frac{A_{\rm hs}}{kpc^2})  }            
\label{rampressure.eqn_2}
\end{equation}
Using the spectral ages (21.9 and 32.6 Myr for the NE and SW lobes) of the
lobes from Table~9 of Jamrozy et al. (2008), assuming the hotspot
diameter to be 10 kpc (Jeyakumar \& Saikia 2000) and measuring
$U_{\rm tot}$ and $n_{\rm p}$ from our radio and X-ray observations, we obtain
the jet-head (or lobe-head) velocities ($\beta_{\rm head}$) to be 0.0084c
and 0.0073c for the NE and SW lobes respectively. The average separation
velocities ($\beta_{\rm sep}$) between the lobe-head and the back-flowing
plasma estimated from the spectral ageing analysis are 0.07c and 0.04c
for the NE and SW lobes respectively (Jamrozy et al. 2008). Since in
general there will be a back-flow velocity ($\beta_{\rm bf}$) of the
relativistic plasma, we can write
\begin{equation}
\beta_{\rm head} + \beta_{\rm bf} = \beta_{\rm sep}
\label{sep-bf_velocity}
\end{equation}
Using Equation~\ref{sep-bf_velocity}, the values of $\beta_{\rm bf}$
come out to be 0.062c and 0.033c for the NE and SW lobes
respectively, which are much higher than the lobe-head velocities.
Since $\beta_{\rm head}$ is estimated from the ram pressure balance at
the jet head, it can be interpreted as the present lobe-head velocity.
Just to compare with the velocity of sound ($v_s$) of the external
medium we have calculated the sound velocity at temperature 2.62 keV,
finding that $v_s=$0.002c. So, the present lobe-head velocities for
both the lobes are supersonic. It should be borne in mind that there
is an ambiguity about the value of $A_{\rm hs}$. Therefore, it is
worth trying to get two limiting values of the diameters corresponding
to two limiting values of $A_{hs}$. If it is the primary hotspot area
that is relevant, then our measurement on the bright hotspot of SW
lobe in the high resolution image of Gilbert et al. (2004) yields a
major axis of $\sim$ 2.7 kpc. Since the image is convolved with a
beam size of 0.25 arcsec ($\sim$1.39 kpc), the deconvolved major axis
(which we assume to be the diameter) is $\sim$2 kpc, which is a good
estimate of the diameter corresponding to the lower limit of $A_{\rm
  hs}$. For the lower limit of $A_{hs}$, the values of $\beta_{\rm
  head}$ are 0.042c and 0.036c for the NE and SW lobes respectively.
In this case, the back-flow velocity of the relativistic plasma will
be 0.028c for NE lobe and 0.004c for the SW lobe. However, the upper
limit of the quantity $A_{\rm hs}$ is not so clear. If the jet moves
about at the end of the lobe on a short timescale, then it might be
appropriate to use the cross-sectional area of the entire front part
of the lobe for $A_{\rm hs}$. From the high resolution map of Gilbert
et al. (2004), our estimation of the entire cross-sectional area
encompassing the region of primary hotspot and two secondary hotspots
in SW lobe yields a diameter of 25 kpc. If this value corresponds to
the diameter of the upper limit of $A_{\rm hs}$, then the values of
$\beta_{\rm head}$ are 0.003c for both the lobes, which are very close
to the sound speed (0.002c) of the 2.62 keV gaseous environment. In
this case, the back-flow velocity of the relativistic plasma has to be
very high compared to the lobe-head velocity.

For our calculations in this section, we have assumed that the lobes
are in the plane of the sky, i.e., at 90$^{\circ}$ to the LOS.
However, the lobes can be inclined at any angle between
45$-$90$^{\circ}$ with the LOS, as 3C\,457 is a radio galaxy rather than
a quasar. It is therefore important to discuss the effect on our
calculation if the lobes are inclined at an angle of 45$^{\circ}$ with
the LOS. The values of the lobe pressure are 0.171$\times$10$^{-12}$
and 0.274$\times$10$^{-12}$ Pa for the NE and SW lobes respectively.
The error bars of the internal pressures in this case are quite
similar to the ones obtained with the assumption that the lobes are in
the plane of the sky. These values are lower by about 24 per cent and
28 per cent respectively than the values obtained with the assumption
that the lobes are in the plane of the sky. If the lobes are at
45$^{\circ}$ with the LOS, then the ratios of the internal to external
pressures at the lobe-head are 1.04$^{+0.61}_{-0.36}$ and
1.50$^{+0.84}_{-0.53}$ for the NE and SW lobes respectively. So the lobe
heads are still approximately at pressure balance. The ratios of
external to internal pressures at the mid points of the NE and
SW lobe are 2.10$^{+1.06}_{-0.62}$ and 1.42$^{+0.69}_{-0.44}$
respectively. So, if the lobes are inclined at an angle of 45 degree
with the LOS, the mid point of NE lobe is still clearly
over-pressured. However, the mid point of SW lobe is close to pressure
balance. In the case of the lobes inclined at 45$^{\circ}$ with the
LOS, the lobe-head velocities (from Equation~\ref{rampressure.eqn_2})
are 0.0088c and 0.0074c in the NE and SW lobes respectively for
$A_{hs}=$10 kpc, which are very close to the values obtained with the
assumption that the lobes are in the plane of the sky. The values of
$\beta_{head}$ are also not significantly altered for the two limiting
values of $A_{hs}$, compared to the values obtained for the lobes in
the plane of ths sky.

Therefore, the qualitative results on the lobe-head velocity and
pressure balance between the lobes and the environment are all more or
less similar irrespective of whether the lobes are in the plane of the
sky or inclined at an angle of 45$^{\circ}$ with LOS.

We now discuss the lateral expansion of the lobes. The
under-pressured parts of the lobes indicate that the lateral expansion
has been subsonic and might have reached a stage when it is
under-pressured. These results are all consistent with the results of
Croston et al. (2004) who have studied two large radio galaxies of
similar kind. This implies that the IC-CMB model for the X-ray
emission from the lobes of 3C\,457 appears to be self-consistent (see
also Croston et al. 2004, 2005). Had we found an internal pressure
much higher than the external pressure of the lobes in the IC-CMB
model with the assumptions adopted for our calculation, we would have
expected at least some contribution to the lobe X-ray emission from
shocked thermal emission. We know from spectral ageing analysis, GRSs
(or large radio galaxies) are older (Jamrozy et al. 2008) and likely
to represent the late stages of evolution of radio galaxies.
Therefore, the results on 3C\,223 and 3C\,284 from Croston et al.
(2004), and on 3C\,457 from the present paper, suggest that we do not
see any evidence of over-pressured lobes at the late stages of
evolution of radio galaxies. In other words, the results on all three
radio galaxies suggest that GRSs (or large radio galaxies) are likely
to be at the late stages of evolution as they are no longer
significantly over-pressured. We would expect over-pressured lobes to
be generally found in smaller radio galaxies, as required for
dynamical models of radio galaxies.

It should be borne in mind that, in carrying out the pressure balance
calculation in the IC-CMB model, we assumed no contribution of protons
in the lobes. With this assumption, the lobe heads are approximately
at pressure balance and the mid-points of the lobes are
under-pressured by a factor of $\sim$1.4 to 2.5 depending upon whether
the lobes are in the plane of the sky or inclined at an angle of
45$^{\circ}$ with LOS. It is interesting to note that if the energy
density in protons were comparable to that of electrons the lobes
would still be under-pressured towards the core and would be in rough
pressure balance at the mid points of the lobes. So a modest
contribution from protons and other non-radiating particles would
somewhat alter our conclusion about pressure balance. For the
supersonic lateral expansion of the lobes, either they have to be
magnetically dominated or dominated by non radiating particles. A more
detailed discussion on supersonic lateral expansion is found in
Section~4.5 of Croston et al. (2004). There is no strong evidence in
the literature for assuming an appreciable amount of non-radiating
particles inside the lobes which can cause supersonic lateral
expansion. In this regard, lobe-related depolarisation measurements of
3C\,457 would be important.

In Section~\ref{Lobe-emission}, we described why we have favoured
IC-CMB model. However, it is important to bear in mind that we were
unable to draw a firm conclusion about the X-ray emission mechanism
based on the spectral fits alone. The pressure balance arguments in
this section are based on assuming the IC-CMB model, which is
self-consistent. If we assume that the lobe-related X-ray emission is
thermal then we have no direct information about the internal
pressure; however, we will have an upper limit on the pressure due to
the electrons only. For supersonic expansion of lobes (as is required
for the thermal interpretation of the lobe-related X-ray emission), we
would require $U_{\rm p}$ or $U_{\rm B} \gg U_{e^{+}e^{-}}$, where
$U_{\rm p}$ is proton energy, $U_{e^{+}e^{-}}$ is the energy of
radiating particles and $U_{\rm B}$ is the magnetic field energy.

%%%%%%%%%%%%%%%%%%%%%%%%%%%%%%%%%%%%%%%%%%%%%%%%%%%%%%%%%%%%%%%%%%%%%%%%%

%%%%%%%%%%%%%%%%%%%%%%%%%%%%%%%%%%%%%%%%%%%%%%%%%%%%%%%%%%%%%%%%%%%%%%%%%

\subsection{Choice of a different value of Galactic hydrogen column density}
\label{choice.of.diff.N_H}
As has been described in Section~\ref{xray.data}, our data suggest a
range of values of $N_{\rm H}$ (0.15$\times$10$^{22}$ cm$^{-2}$ $\le$ $N_{\rm H}$
$\le$ 0.225$\times$10$^{22}$ cm$^{-2}$ ). Any value below
0.15$\times$10$^{22}$ cm$^{-2}$ worsens the quality and feasibility of
the fit to all the spectra, and there is no strong evidence for any
significantly higher value than 0.225$\times$10$^{22}$ cm$^{-2}$ . The
results discussed above assume $N_{\rm H}=$0.225$\times$10$^{22}$ cm$^{-2}$.
We will now discuss how our results are affected if we assume
$N_{\rm H}= 0.15 \times 10^{22}$ cm$^{-2}$ (hereafter referred to as
the `lower
value of $N_{\rm H}$') instead of $N_{\rm H}=0.225 \times 10^{22}$ cm$^{-2}$
(hereafter referred to as the `higher value of $N_{\rm H}$).
We have provided the best fit values of all the parameters of all the spectra 
for $N_{\rm H}=$0.15$\times$10$^{22}$ cm$^{-2}$ in Tables~\ref{fitstat_core_nH=0.15} 
and ~\ref{fitstat_slobe.n.env_nH=0.15}.

For the X-ray spectrum of the core of 3C\,457, in Model I, the 1 keV
X-ray flux density of the soft power law is 3.84 nJy for the lower
value of $N_{\rm H}$. This is  a little lower than obtained for the higher
value of $N_{\rm H}$; however, 3.84 nJy is still consistent with the X-ray
and radio flux density correlation as found by Hardcastle \& Worrall
(1999). In Model II if we use the lower value of $N_{\rm H}$, the soft mekal
component has a best-fitting temperature of 3.99 keV, a little higher than
that obtained for the higher value of $N_{\rm H}$, whereas, the soft mekal
flux density has similar value and is not well constrained on the
upper side for the lower value of $N_{\rm H}$, as is the case for the higher
value. The best-fitting values of all the other fit parameters for both
Model I and II of the core X-ray spectrum are similar within the
errors. The fit quality is equally good for both values of $N_{\rm H}$.

For the X-ray spectrum of the SW lobe, if we consider a power law then
both the parameters (photon index and 1 keV flux density) have similar
best-fitting values within the errors. If we consider the mekal model fit
to the SW lobe, then the best-fitting temperature is even higher and not
well constrained both above and below; however, the unabsorbed flux of
the mekal model is similar within the errors. So our qualitative
results and reasons for favouring the power-law model, not the mekal
model, still hold. With the lower value of $N_{\rm H}$ the magnetic fields
are 0.77 and 0.45 nT for the NE and SW lobes respectively. The
corresponding total pressures are 0.209$\times$10$^{-12}$ and
0.317$\times$10$^{-12}$ Pa. The error bars shown for magnetic fields and
pressures are similar to those estimated for the higher value of $N_{\rm H}$.
Ultimately, our results that the lobe-heads are close to pressure
balance and the mid points of the lobes are under-pressured are still
valid.

For the environments around 3C\,457, again the fits are quite similar
within the errors for the lower $N_{\rm H}$ value. The best-fitting value of the
mekal temperature is 3.42$^{+1.83}_{-1.00}$ keV. The bolometric
luminosity (0.1$-$100 kev), corrected for the masked regions around
the lobes and core, is $\sim$1.38$^{+1.12}_{-0.87}$ $\times$10$^{44}$
erg s$^{-1}$. At this environmental luminosity, the L$_X$$-$T$_X$
correlation (Equation~\ref{Lx-Tx_corrln}) of the clusters predicts a
temperature of 3.32$^{+1.20}_{-1.10}$ keV. Given the error bars, we
cannot definitely say that there is any heating effect due to the
$P{\rm d}V$ work done by the lobes on to the environment.

In conclusions, all our results are qualitatively similar for either
$N_{\rm H}$ value.

%%%%%%%%%%%%%%%%%%%%%%%%%%%%%%%%%%%%%%%%%%%%%%%%%%%%%%%%%%%%%%%%%%%%%%%%%%%%%%%%%%%%%%%%%%%%%%%%%%%%%%%%%%%%%%%%%
%\vspace{-10cm}
\begin{table*}
\caption{Fit statistics of the spectrum of the core with $N_{\rm H}=$0.15$\times$10$^{22}$ cm$^{-2}$   }
\label{core_w.N_H=0.15}
\begin{tabular}{llllccc}
\hline\hline
Model          & Parameter                 & Model I                        &    Model II           \\
component      &                           &                                &                       \\
  (1)          &  (2)                      & (3)                            &     (4)                \\
\hline
%------------------------------------------------------------------------------------------------------------
Soft power law & $\Gamma$                  & 1.81$^{+0.28}_{-0.34}$         &                        \\
               & 1 keV flux density        & 3.84$^{+0.33}_{-0.33}$         &                        \\
               & (nJy)                     &                                &                        \\\hline
%------------------------------------------------------------------------------------------------------------
Soft mekal     & kT (keV)                  &                                & 3.99$^{+4.65}_{-1.57}$  \\
               & Unabsorbed flux           &                           &(2.75$^{+26.70}_{-0.73}$)$\times10^{-14}$ \\
               & (erg/s/cm$^2$)            &                                &                      \\ \hline
%------------------------------------------------------------------------------------------------------------
Hard power law & Nuclear $N_{\rm H}$ (cm$^{-2}$) &(29.7$^{+9.8}_{-4.5}$)$\times10^{22}$ & (29.0$^{+8.9}_{-4.5}$)$\times10^{22}$ \\
               & $\Gamma$                  & 1.53$^{+0.16}_{-0.07}$         & 1.50$^{+0.08}_{-0.07}$                \\
               & Unabsorbed flux           &(9.95$^{+2.53}_{-9.87}$)$\times10^{-13}$ &(9.84$^{+2.53}_{-9.49}$)$\times10^{-13}$   \\
               & (erg/s/cm$^2$)            &                                        &                       \\ \hline
$\chi_{red}^2$ &                           & 0.66                                   &  0.67                 \\
%------------------------------------------------------------------------------------------------------------
\hline
\end{tabular} \\
Note: The observed part of the spectrum has been integrated to find the flux.
\label{fitstat_core_nH=0.15}
\end{table*}

%%%%%%%%%%%%%%%%%%%%%%%%%%%%%%%%%%%

\begin{table*}
\caption{Fit statistics of the spectrum of SW lobe and environment with $N_{\rm H}=$0.15$\times$10$^{22}$ cm$^{-2}$ }
\label{slobe_w.N_H=0.15}
\begin{tabular}{llllccc}
\hline\hline
 Source              & Model          & Parameter              & Best
 fit values    \\
 Component           &                &                        &                     \\
  (1)                &  (2)           & (3)                    &  (4)               \\
\hline
%------------------------------------------------------------------------------------------------
  SW lobe             & Power law      & $\Gamma$               & 1.41$^{+0.20}_{-0.20}$   \\
                     &                & 1 keV flux density     & 3.17$^{+0.38}_{-0.39}$   \\
                     &                &   (nJy)                &                          \\
                     &                & $\chi_{red}^2$         & 0.80                     \\
%------------------------------------------------------------------------------------------------
                     & mekal          & kT (keV)               & 18.58$^{+61.32}_{-10.96}$  \\
                     &                & Unabsorbed flux        &(3.33$^{+0.82}_{-3.28}$)$\times 10^{-14}$ \\
                     &                & (erg/s/cm$^2$)         &                          \\
                     &                &$\chi_{red}^2$          & 0.83                     \\ \hline
%------------------------------------------------------------------------------------------------
 Environment         & mekal          & kT (keV)               & 3.42$^{+1.83}_{-1.00}$        \\
                     &                & Unabsorbed flux        &(6.00$^{+1.21}_{-1.47}$)$\times 10^{-14}$  \\
                     &                & (erg/s/cm$^2$)         &                             \\
                     &                &$\chi_{red}^2$          & 1.20                        \\
\hline
\end{tabular} \\
Note: The observed part of the spectrum has been integrated to find the flux.
\label{fitstat_slobe.n.env_nH=0.15}
\end{table*}

%%%%%%%%%%%%%%%%%%%%%%%%%%%%%%%%%%%%%%%%%%%%%%%%%%%%%%%%%%%%%%%%%%%%%%%%%%%%%%%%%%%%%%%%%%%%%%%%%%%%%%%%%%%%%%

\section{Concluding remarks}
We have presented multi-frequency radio observations with the GMRT and
VLA and X-ray observations with {\it XMM-Newton} of a giant radio
galaxy, 3C\,457. At radio wavelengths we have detected the core and
lobes, while at X-ray wavelengths we have detected the core, lobes
and thermal gaseous environment around the source. Our multi-frequency
radio observations have allowed us to constrain the radio spectra of
the components within the observed frequency range, and we found that
the spectra of all components as well as the integrated spectrum are
consistent with a power law. The X-ray spectrum of the core is well
fitted by a composite spectrum consisting of a hard heavily absorbed
power law and a soft unabsorbed power law. The hard absorbed power law
is a characteristic of an AGN with the absorption due to the torus.
The soft unabsorbed power law is assumed to originate from the jet
base. It is difficult to know the origin of this emission: it could be
due to synchrotron, inverse-Compton, or a mixture of both, which
cannot be disentangled with our data.

We have shown that an IC-CMB model for the X-ray emission from the lobes is
self-consistent. For both the lobes, the magnetic field determined
from IC-CMB modelling is within a factor of two of the minimum energy
value. We find that both the lobes are close to pressure balance near
the hotspots but under-pressured towards the core, if we neglect the
contribution from any sort of non-radiating particles inside the
lobes. If the energy density in protons were comparable to that of
electrons the lobes would still be under-pressured towards the core
and would be in rough pressure balance at the mid points of the lobes.
So a modest contribution from protons and other non-radiating
particles would alter our conclusion about pressure balance. If the
lobes have supersonic lateral expansion, then either they have to be
magnetically dominated or dominated by non radiating particles. There
is no strong evidence in the literature for assuming an appreciable
amount of non-radiating particles inside the lobes. A model in which
there is no supersonic lateral expansion is consistent with our
interpretation of the X-ray emission as originating in IC-CMB
emission, rather than shocked hot gas surrounding the lobes.

A poor cluster scale environment around the source has been detected
unambiguously, with a best-fitting temperature of
2.62$^{+1.15}_{-0.69}$keV. The environment is consistent with an
isothermal gas with a $\beta$ model profile (e.g. Birkinshaw \&
Worrall 1993). The environment of 3C\,457 follows the $L_{\rm
  X}$-$T_{\rm X}$ correlation for clusters, which means that we can
identify no significant global heating effect due to $P{\rm d}V$ work on to
the environment. The number of counts is insufficient to check whether
there are any temperature variations in the atmosphere.

The results we present here for 3C\,457, together with the results on
3C\,223 and 3C\,284 from Croston et al. (2004), suggest that there is no
strong evidence for over-pressured lobes in large radio galaxies or
GRSs, i.e. that GRSs (or large radio galaxies) are likely to be in the
late stages of evolution as they are no longer significantly
over-pressured.

\section*{Acknowledgments}
This work is partly based on observations obtained with {\it
XMM-Newton}, an ESA science mission with instruments and contributions
directly funded by ESA Member States and the USA (NASA).
The Giant Metrewave Radio Telescope is a national facility operated by
the National Centre for Radio Astrophysics of the Tata Institute of
Fundamental Research. The National Radio Astronomy Observatory is a
facility of the National Science Foundation operated under
co-operative agreement by Associated Universities Inc. This research
has made use of the NASA/IPAC extragalactic database (NED) which is
operated by the Jet Propulsion Laboratory, Caltech, under contract
with the National Aeronautics and Space Administration. We thank
numerous contributors to the GNU/Linux group. CK acknowledges travel
support and local hospitality provided, during May 2008, by the University
of Hertfordshire, Hatfield, UK, where this research was partly done.
CK acknowledges Ranjeev Misra and Gulab C. Dewangan from IUCAA for
useful discussions regarding data analysis and interpretations. MJH
thanks the Royal Society for generous funding through the Research
Fellowships scheme.

{}

\end{document}